\documentclass{ws-ijmpe}
\usepackage{amssymb}

\begin{document}

\markboth{Tong \& Xu}{Magnetars: fact or fiction?}

\catchline{}{}{}{}{}

\title{Magnetars: fact or fiction?}

\author{H. Tong}

\address{Institute of High Energy Physics, Chinese Academy of Sciences,\\
Beijing 100049, China; haotong@ihep.ac.cn}

\author{R.~X. Xu}

\address{School of Physics and State Key Laboratory of Nuclear Physics and
Technology,\\
Peking University, Beijing 100871, China; r.x.xu@pku.edu.cn}

\maketitle


\begin{abstract}
Anomalous X-ray pulsars (AXPs) and soft gamma-ray repeaters (SGRs)
are enigmatic pulsar-like objects. The energy budget is the
fundamental problem in their studies. In the magnetar model, they
are supposed to be powered by the extremely strong magnetic fields
($\gtrsim 10^{14}$ G) of neutron stars. Observations for and against
the magnetar model are both summarized. Considering the difficulties
encountered by the magnetar model to comfortably understand more and
more observations, one may doubt that AXPs and SGRs are really
magnetars. If they are not magnetar candidates (including
magnetar-based models), then they must be ``quark star/fallback
disk'' systems.
\end{abstract}

\section{Introduction}

Since the discovery of pulsars in 1967, many kinds of pulsar-like
objects have been discovered. Among them, anomalous X-ray pulsars
(AXPs) and soft gamma-ray repeaters (SGRs) are two sorts of
enigmatic sources\cite{Mereghetti2008}. Most of their persistent
X-ray luminosities are in excess of their rotational energy loss
rates, and they show no binary signature. Furthermore, they also
show recurrent bursts. Some of the bursts (giant flares) are highly
super-Eddington (with luminosity $\sim 10^{42}
\,\mathrm{erg}\,\mathrm{s}^{-1}$ in the pulsating tail). Therefore,
the fundamental problem in AXP and SGR studies is to find a reliable
power and to balance the energy budget for both their persistent and
burst emissions.

In the magnetar model for AXPs and SGRs, the compact stars are
supposed to be powered by extremely strong magnetic
field\cite{TD1995,TD1996}. In addition, the strong surface dipole
field ($>B_{\mathrm{QED}}\equiv 4.4\times 10^{13} \,\mathrm{G}$)
also provides the braking mechanism of AXPs and SGRs\cite{DT92}. The
decay of strong multipole fields ($\sim 10^{14}-10^{15}
\,\mathrm{G}$) powers their persistent
emissions\cite{TD1996,TLK2002,Beloborodov2007}. Possible sudden
release of magnetic energy (e.g., magnetic reconnection) is
responsible for the bursts\cite{TD1995}. The suppression of Thomson
scattering cross section in strong magnetic field may explain the
super-Eddington luminosity\cite{Paczynski1992,TD1995}. However,
there are accumulating challenges to the magnetar model in recent
observations (see Section 3 below). Therefore, alternative modeling
of AXPs and SGRs are not only possible but also very necessary.

\begin{figure}
\label{PPdot}
 \includegraphics{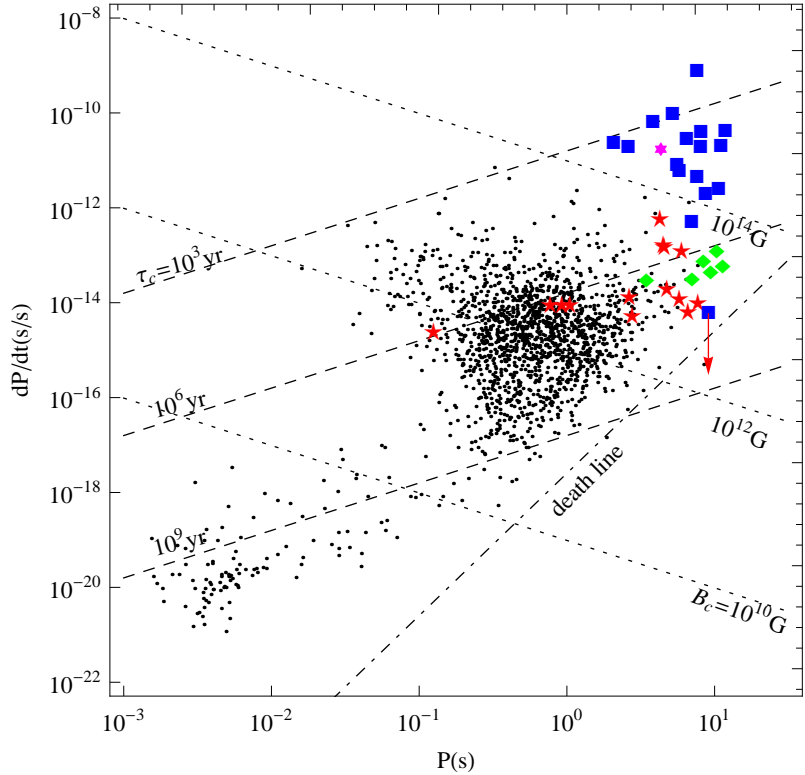}
\caption{P-$\mathrm{\dot{P}}$ diagram of pulsars. Squares are for
AXPs and SGRs, the six-pointed-star is for the radio loud magnetar,
the down-arrow marks the low magnetic field SGR (from McGill online
catalog:
http://www.physics.mcgill.ca/$\sim$pulsar/magnetar/main.html).
Diamonds are for X-ray dim isolated neutron stars (XDINSs) (from
Kaplan \& van Kerkwijk 2011). Stars are for rotating radio
transients (RRATs), dots are for normal and millisecond
pulsars (from ATNF:
http://www.atnf.csiro.au/research/pulsar/psrcat/). }
\end{figure}

AXPs and SGRs may alternatively be fallback disk systems. Accretion
from a supernova fallback disk provides both the braking mechanism
and the persistent emissions\cite{Alpar2001,Chatterjee2000}. The
period clustering of AXPs and SGRs is a natural consequence of disk
braking, as shown in Fig. \ref{PPdot}. The super-Eddington bursts
could be due to the presence of a bare quark
surface\cite{Alcock1986a,Alcock1986b} if the compact star is a quark
star\footnote{The glitch problem for strange stars raised by
Aplar\cite{Alpar1987} is solved in the solid quark star
domain\cite{Xu2003b,Zhou2004}. Cold quark matter is suggested in
quark-clustering state there and the star behaves then like a solid
star with rigidity.}. The energy of bursts may be from elastic and
gravitational energy release during star quakes\cite{Xu2006,Xu2007}.
Therefore, AXPs and SGRs could be ``quark star/fallback disk''
systems. In this scenario, only normal strength magnetic field is
required $\sim 10^{12} \,\mathrm{G}$, the same as that of normal
radio pulsar.

Observations for and against the magnetar model are summarized in
sections 2 and 3, respectively. Then in section 4, various
alternative modelings of AXPs and SGRs are discussed. In section 5,
we try to answer the question: what if no magnetars exists at all?
Our conclusions are given in section 6: \emph{AXPs and SGRs may be
magnetars. However, it is also possible that they are not magnetars.
If AXPs and SGRs are not magnetars (including magnetar-based
models), then they must be ``quark star/fallback disk'' systems.}

\section{Observations for the magnetar model}

In the following we will summarize observations for the magnetar
model\cite{Mereghetti2008}. The limitations of these observations
are also presented.

\begin{enumerate}
 \item
Measurement of magnetic field through period and period
derivative is often taken as confirmation of a
magnetar\cite{Kouveliotou1998}. This assumes that AXPs and SGRs are
braked down by magnetic dipole radiation like that of
rotation powered pulsars. However, if AXPs and SGRs are magnetic
energy powered, then a strong particle wind will also contribute to
the braking torque. The corresponding dipole field is no longer so
high\cite{Harding1999}. In the fallback disk model, AXPs and SGRs
are braked down by propeller
effect\cite{Alpar2001,Chatterjee2000} and only normal strength
magnetic field ($\sim 10^{12} \,\mathrm{G}$) is required. More
importantly, in the magnetar model, the dipole field mainly provides
the braking torque. It is the multipole field that powers
the star's persistent and burst emissions in the magnetar
model\cite{TD1995,TD1996}. Therefore, magnetic field calculated from
period and period derivatives can not be taken as confirmation of a
magnetar.

\item
The strength of multipole field may be measured from spectral lines.
Possible discovery of cyclotron lines during outburst is
claimed\cite{Ibrahim2002}. However, the observations are not
conclusive. Furthermore, whether they are due to proton or electron
cyclotron lines are not certain. In the future, if strong electron
cyclotron lines are found from AXPs/SGRs (e.g., like that of 1E
1207.4-5209\cite{Xu2003a,GH07}), then AXPs and SGRs must have normal
strength magnetic field. Meanwhile, the absence of atomic
features in AXP and SGR persistent emissions may imply that they are
quark stars, like that of X-ray dim isolated neutron stars
(XDINSs)\cite{Xu2002}.

\item
The pulsating tail seen in SGR giant flares requires a strong
confinement magnetic field. In order to confine a fireball with
energy $\sim 10^{44}\,\mathrm{erg}$, a magnetic field higher than
$10^{14}\,\mathrm{G}$ is required. However, this assumes that the
energy is release suddenly. If the energy release is in a continuous
process, the requirement on magnetic field strength is no longer
available\cite{Katz1996,Zhang2000}.

\item
The super-Eddington luminosity in the pulsating tail of SGR giant
flares is due to magnetic suppression of Thomson scattering cross
section. This point is one of the key arguments to introduce
magnetar strength magnetic field\cite{Paczynski1992,TD1995}.
However, this consideration neglects other force terms (e.g.,
magnetic stress). When this effect is included, only normal strength
magnetic field is required\cite{Spruit2010}. On the other hand,
AXPs and SGRs may be quark stars. The presence of a bare quark
surface will explain the super-Eddington luminosity
naturally\cite{Alcock1986a,Alcock1986b}. Not only the
super-Eddington luminosity in the pulsating tail but also that in
the initial spike are allowed if AXPs and SGRs are {\em bare} quark
stars.

\item
The SGR-like burst seen in the high magnetic field pulsar PSR
J1846-0258 ($B_{\mathrm{dip}}=4.9\times 10^{13}\,\mathrm{G}$) is for
the magnetar model\cite{Gavriil2008}. This point is a little
misleading, since there is also a low magnetic field SGR
0418+5729\cite{Rea2010} with dipole magnetic field smaller than
$7.5\times 10^{12}\,\mathrm{G}$. The corresponding explanation in
the magnetar model is that it is the multipole field (not the dipole
field) which is responsible for the star's bursts and persistent
emissions. If we assume that AXPs and SGRs are ``quark
star/fallback disk'' systems, their bursts are due to energy
release during accretion induced star quakes (AIQs)\cite{Xu2006}. If
the same mechanism can occur also in PSR J1846-0258, SGR-like bursts
are available. The glitch\cite{Livingstone2010} detected
in PSR J1846-0258 is consistent with the AIQ scenario.

\item
There are also other observations which may point to a strong
magnetic field, e.g., energy of persistent emissions, energy of
bursts, spectral modeling, etc. However, these arguments are more
model-dependent than the points listed above.

\end{enumerate}

In summary, there are two assumptions in the magentar model: a
strong dipole field and a strong multipole field. Under these two
assumptions, the magnetar model can explain many aspects of AXP and
SGR's observations. However, up to now, we have no direct and clear
evidence that AXPs and SGRs have super-strong magnetic field.
Meanwhile, if we assume that AXPs and SGRs are ``quark star/fallback
disk'' systems, these observations can be explained equally well.
The period clustering and super-Eddington luminosity are natural
consequences in the ``quark star$+$fallback disk'' model.

\section{Failed predictions: challenges to the magnetar model}

There are accumulating difficulties for the magnetar model in recent observations.
Below, we will give several examples.

\begin{enumerate}

 \item The magnetars are assumed to be born with a very short
rotation period $\sim 1 \,\mathrm{ms}$. This will result in a very
high magnetic field ($\sim 10^{14}-10^{15} \,\mathrm{G}$) through
the dynamo process. Meanwhile, since there are more rotational
energy and the magnetic field is very high, the corresponding
supernova will be more energetic and the putative magnetar will have
a very large kick velocity\cite{DT92}. However, these two
predictions are both rejected by later observations.
It is found that the supernova explosion energy are of normal value
by analyzing supernova remnants associated with AXPs and
SGRs\cite{Vink2006}. Their surrounding environment is not different
from that of normal pulsars\cite{Gaensler2001}. Therefore, from the
supernova explosion point of view, we see no difference between
AXPs/SGRs and normal pulsars.

 \item Additionally, proper motion measurement of the radio emitting
magnetar AXP XTE J1810-197 gives even a smaller value than that of
normal pulsars\cite{Helfand2007}. A low kick velocity is also
required by studying the supernova remnants associated with AXPs and
SGRs\cite{Gaensler2001}. Therefore, from the kick velocity point of
view, we see no difference between AXPs/SGRs and normal pulsars.

 \item In the magnetar model, it is commonly assumed that AXPs and
SGRs are braked down by magnetic dipole radiation. Therefore, they
have very strong surface dipole field\cite{Kouveliotou1998}. The
magnetic dipole braking means that the rotation energy of AXPs and
SGRs are taken away by similar processes to that of rotation powered
pulsars. Therefore, we should see some rotation powered activities
in AXPs and SGRs if there are really magnetars\cite{ZhangBing2003}.
When applying the outer gap model to AXPs and SGRs, it was predicted
that they will emit high-energy gamma-rays which are detectable by
Fermi-LAT\cite{Cheng2001}. However, no significant detection is
reported for all AXPs and SGRs in Fermi-LAT
observations\cite{Mus2010,Abdo2010}. It is shown that there are
conflicts between the prediction of the outer gap model in the case
of magnetars and Fermi-LAT observations\cite{Tong2010,Tong2011}.
Fermi-LAT observations tell us that AXPs and SGRs must be either
magentars without strong surface dipole field or fallback disk
systems. The detection of high-energy gamma-ray emissions from one
high magnetic field pulsar is for the above
analysis\cite{Parent2011}.

 \item The traditional picture about magnetars is that: they are young neutron stars;
they have super-strong surface dipole field; their multipole field
is as high as the dipole field. However, this picture is challenged
by the low magnetic field SGR 0418+5729\cite{Rea2010}. It has a
rotation period $P=9.08\,\mathrm{s}$ and a period derivative
$\dot{P}<6.0\times 10^{-15}$. The implied surface dipole magnetic
field is less than $7.5\times 10^{12} \,\mathrm{G}$, and a
characteristic age larger than $2.4\times 10^7 \,\mathrm{yr}$.
Therefore, the traditional picture about magnetars does not apply to
this source. In order to power its persistent and burst emissions, a
magnetar strength of multipole field is {\em merely}
assumed\cite{Rea2010}. Whether a dynamo process can generate such a
field configuration is not certain. Moreover, if people assume that
such an aged magnetar (with age $\sim 10^6-10^7 \,\mathrm{yr}$) is
still burst-active\cite{Turolla2011}, there will be too many SGRs in
our Galaxy\cite{Muno2008}. The seven XDINSs could be high magnetic
field neutron stars ($B_{\mathrm{dip}}\sim 10^{13} \,\mathrm{G}$)
with age $\sim 10^6 \,\mathrm{yr}$ in the magneto-dipole braking
scenario. If a magnetar at the age of SGR 0418+5729 can still be
burst-active, then we should also have detected some SGR-like
activities in XDINSs. However, such activities have never been
observed\cite{Tong2010ins,Tong2011ins}.

 \item The radio variability of PSR J1622-4950 is assumed to be due its magnetar nature\cite{Levin2010}
(a ``radio loud'' magnetar in X-ray quiescence).
Its X-ray luminosity is much smaller than other typical AXP/SGR's X-ray luminosity.
Why it has such a low X-ray luminosity? The same question also applies to the transient
magnetars\cite{Kaspi2007}. How can a magnetar strength field decay
in one source and not decay in another? How can it decay some times and not decay some other times?

\end{enumerate}

In conclusion, these observations provide challenges to the magnetar
model. We surely require alternative ideas to understand the
behaviors of AXPs and SGRs. There may be alternative origins for
strong magnetic field\cite{Peng2007}. Under the general assumption
of magnetic energy powered, alternative modeling of AXP and SGR's
persistent and burst emissions can be done. People can even build
models for AXPs and SGRs without the inclusion of any magnetar
strength field.

\section{Alternative modelings of AXPs and SGRs}

From the above analyses, we can get the conclusion that AXPs and
SGRs may be magnetars. However, it is also possible that they are
actually not magnetars. If AXPs and SGRs are not magnetars, then we
must find alternative models to reproduce the general results
observed for AXPs and SGRs. In fact, there do exist various
alternative modelings of AXPs and SGRs.

\begin{enumerate}

 \item It is possible that AXPs and SGRs are of wind braking. A
strong multipole field provides both the persistent and burst
emissions of AXPs and SGRs. A particle wind which originates from
decay of strong multipole field could contribute significant braking
torque\cite{Harding1999}.

 \item A fallback disk may coexist with a magnetar strength
multipole field\cite{Ertan2007}. In this magnetar-based hybrid
model, the braking and persistent emissions of AXPs and SGRs are
provided by the fallback disks. While the bursts (especially giant
flares) are powered by the strong multipole field, the same as the
magnetar case.

 \item Note that the super-Eddington bursts of SGRs can be
explained naturally in the quark star model because of
self-bound\cite{Alcock1986a}, it is possible that AXPs and SGRs are
quark stars instead of being normal neutron stars. The bursts and
giant flares may be from energy release during star quakes (AIQ
model\cite{Xu2006}) of solid quark stars. At the same time, a
fallback disk provides the braking and persistent emissions of AXPs
and SGRs. This ``quark star+fallback disk'' scenario is discussed
with details in the following section.

 \item The color interaction between quarks in quark matter may be
stronger than that in normal nucleon matter. Therefore, magnetars
could be strongly magnetized quark stars instead of strongly
magnetized neutron stars. A strongly magnetized quark star
surrounded by a degenerate quark nova remnant may explain the
observations of AXPs and SGRs\cite{Ouyed2011}.

 \item If the compact star is a massive white dwarf, due to a
larger momentum of inertia, its rotational energy is enough to power
the emissions of AXPs and SGRs. A massive white dwarf surrounded by
a fossil disk may provide an alternative modeling of AXPs and
SGRs\cite{Malheiro2011}.

\end{enumerate}

In the above list of discussions, models (1) and (2) are
magentar-based ones. Models (3) and (4) involve different kinds of
quark stars, in either solid or liquid states. Models (1)-(4) are
all in the neutron star domain (normal neutron stars or quark
stars), while model (5) is based on white dwarfs. Multiwave
observations of AXPs and SGRs may help us to finally distinguish
between those different models in the future.

\section{What if no magnetars exists at all?}

Now, we will outline how AXPs and SGRs can be modeled in the
``quark star$+$fallback disk'' scenario.

\begin{enumerate}

 \item Spindown and persistent emissions.
Neutron stars (including normal neutron stars and quark stars) are born in supernova explosions.
Some of the explosive material may fallback onto the neutron star. If the fallback material
carries some amount of angular momentum, they may form a disk, i.e., supernova fallback disks.
Considering the period clustering and persistent emissions of AXPs and SGRs, fallback disk model
for them are proposed\cite{Alpar2001,Chatterjee2000}. The equilibrium period is reached
when the corotation radius equals the magnetospheric radius\cite{Frank2003}:
\begin{equation}
 P_{\mathrm{eq}}= 8\,\mathrm{s} \,\mu_{30}^{6/7} M_1^{-5/7} \dot{M}_{15}^{-3/7}.
\end{equation}
This can explain the period clustering of AXPs and SGRs naturally. In the fallback scenario,
a large period derivative is not required. The low magnetic field
SGR 0418+5729 is such an example. Although its period derivative is very small, its period
is the same as other AXPs and SGRs. This suggests that it must also be braked down by a fallback
disk\cite{Alpar2011}. Now its period derivative is very small since the disk lies outside
the light cylinder and has little interaction with the central star.

The persistent emission spectral of AXPs and SGRs can be modeled similarly to that of
accretion systems. It is shown that both the soft X-ray and hard
X-ray spectrum of AXP 4U 0142+61 can be modeled uniformly employing
the bulk motion Comptonization process\cite{Trumper2010}. The
discovery of a debris disk around AXP 4U 0142+61 is for the fallback
disk model\cite{Wang2006}. If a massive neutron star (including
normal neutron stars and quark stars) has a fallback disk it will be
an AXP/SGR\cite{Xu2007}. Otherwise, it will be a normal pulsar. This
is why we discovery a debris disk only around an AXP not in other
young normal pulsars. The optical and IR emission of this source is
consistent with a gaseous accretion disk\cite{Ertan2007}.

\item Super-Eddington luminosity.
Whether pulsars are normal neutron stars (mainly composed of nucleon
matter) or quark stars (composed of deconfined quark matter) is a
fundamental problem in pulsar astrophysics\cite{Xu2011}. When the
quark star model for pulsars was proposed, it was noted that the
``very high luminosity event'' of 1979 March 5 may imply the
existence of a bare quark star surface\cite{Alcock1986a}. Because of
the bare quark star surface, both the super-Eddington radiations in
the pulsating tail and in the initial spike are
allowed\cite{Alcock1986b}. The existence of a bare quark star
surface may also help to explain other bursting phenomena, e.g.,
supernova explosions\cite{Chen2007} and gamm-ray
bursts\cite{Paczynski2005}.

\item Energy of bursts (including giant flares).
In the magnetar model, the main observational manifestations (burst and persistent emissions)
are due to magnetic energy.
Meanwhile, the rotational energy and the elastic energy are also present. In the ``quark
star$+$fallback disk'' model, the persistent emissions of AXPs and SGRs are due to
accretion power. The bursts are due to sudden energy release of the quark star,
which may include elastic energy, gravitational energy, and conversion energy from
normal matter to quark matter. The collision by comet-like objects are proposed
for giant flares of SGRs\cite{Alcock1986b,Zhang2000}. The corresponding times scales
and pulse profiles are consistent with observations\cite{Cheng1998,Usov2001}.
Noting the possible connection of AXP/SGR bursts and glitches, an AIQ model
for AXP/SGR bursts is proposed\cite{Xu2006,Lai2009}. The gravitational
energy release during star quakes can be estimated as\cite{Xu2006}
\begin{equation}
 \Delta E =\frac{G M^2}{R} \left|\frac{\Delta R}{R}\right| \sim 5\times 10^{47} \,\mathrm{erg} \,\frac{|\Delta R/R|}{10^{-6}}.
\end{equation}
Therefore, during a glitch with amplitude $\Delta \nu/\nu=2\times 10^{-6}$, a maximum
amount of energy $\sim 5\times 10^{47} \,\mathrm{erg}$ can be released. This is enough to power the
SGR giant flares, including photon energy, neutrinos, and gravitational waves, etc.
The glitch associated with outburst of AXP 1E 2259+586 is consistent with the AIQ model\cite{Kaspi2003}.

Considering the trigger of star quakes, there may be two kinds of
glitches in AXPs and SGRs. One kind of star quake is triggered by
stress build up during the spin down of the star. The time scale for
stress build up may be relatively long, and the star quake may occur
deep inside the star. This kind of star quake will mainly result in
transfer of angular momentum, i.e. glitches. Since the star quake
happens deep inside the star, it will not accompanied by SGR-type
bursts. Some of the AXP/SGR glitches and almost all of the glitches
in normal pulsars (e.g., Vela pulsar\cite{Helfand2001}) are of this
kind. We call this kind of glitches ``{\em spin down induced
glitches}''. Another kind of star quake is trigger by stress build
up when the accretion matter accumulates on the star surface. The
corresponding time scale for stress build up may be relatively short
and the star quake mainly happens near the star surface. This kind
of star quake will not only spin up the star (i.e., glitches) but
also trigger SGR-type bursts. Since the magnetic field lines are
anchored on the star surface, a star quake near the surface will
twist the magnetic field lines, accelerate particles, thus result in
SGR-type bursts (similar to the corona model of
magnetars\cite{Beloborodov2007}). We call this kind of glitches
``{\em accretion induced glitches}''. Therefore, all glitches in
AXPs and SGRs are not accompanied by bursts. Instead, all bursts
should associated with glitches. Because of a larger timing noise
and the sparse of observations, not all glitches during outburst can
be discovered\cite{Mereghetti2008}. Future theoretical and
observational studies are necessary in this scenario of glitch.

\end{enumerate}

Summarily, both the bursts and persistent emissions of AXPs and SGRs
are understandable in the ``quark star$+$fallback disk'' model. The
period clustering and super-Eddington luminosity bursts are natural
consequences of this model.

\section{Summary}

We have discussed both the magnetar model and the fallback disk
model for AXPs and SGRs in previous sections. There are two
assumptions in the magnetar model: a strong dipole field and a
strong multipole field. However, the origin of strong fields, the
presence of strong dipole field and even the presence of strong
multipole field are challenged by recent observations. When studying
AXPs and SGRs, the magnetic dipole braking is often assumed. Many of
the problems are associated with this assumption. Therefore, the
study of AXP and SGR braking mechanism in the future may help us
make clear some of these problems. It may also help us to
distinguish between the magnetar model and the fallback disk model.

Alternative modeling in the ``quark star$+$fallback disk'' scenario
is helpful to understand the nature of AXPs and SGRs. The discovery
of a debris disk and a low magnetic field SGR have deepened our
understanding of AXPs and SGRs. More observations in the
coming decades (e.g., 10 to 30 years) will tell us clearly whether
they are magnetars or ``quark star/fallback disk'' systems.

\section*{Acknowledgements}

We would like to thank M. Malheiro, R. Ruffini for meaningful
discussions. We would like to acknowledge valuable discussions at
the PKU pulsar group. This work is supported by the National Natural
Science Foundation of China (grants 10935001 and 10973002), the
National Basic Research Program of China (grant 2009CB824800), and
the John Templeton Foundation.


\begin{thebibliography}{99}

\bibitem{Mereghetti2008}
S. Mereghetti, {\it A\&ARv} {\bf 15} (2008) 225.

\bibitem{TD1995}
C. Thompson, R. C. Duncan, {\it MNRAS} {\bf 275} (1995) 255.

\bibitem{TD1996}
C. Thompson, R. C. Duncan, {\it ApJ} {\bf 473} (1996) 322.

\bibitem{DT92}
R. C. Duncan, C. Thompson, {\it ApJ} {\bf 392} (1992) L9.

\bibitem{TLK2002}
C. Thompson, M. Lyutikov, S. R. Kulkarni, {\it ApJ} {\bf 574} (2002) 332.

\bibitem{Beloborodov2007}
A. M. Beloborodov, C. Thompson, {\it ApJ} {\bf 657} (2007) 967.

\bibitem{Paczynski1992}
B. Paczynski, {\it ACTA ASTRONOMICA} {\bf 42} (1992) 145.

\bibitem{Alpar2001}
M. A. Alpar, {\it ApJ} {\bf 554} (2001) 1245.

\bibitem{Chatterjee2000}
P. Chatterjee, L. Hernquist, R. Narayan, {\it ApJ} {\bf 534} (2000) 373.

\bibitem{Kaplan2011}
D. L. Kaplan, M. H. van Kerkwijk, (2011) arXiv:1109.2105.

\bibitem{Alcock1986a}
C. Alcock, E. Farhi, A. Olinto, {\it ApJ} {\bf 310} (1986) 261.

\bibitem{Alcock1986b}
C. Alcock, E. Farhi, A. Olinto, {\it PRL} {\bf 57} (1986) 2088.

\bibitem{Alpar1987}
M. A. Alpar, {\it PRL} {\bf 58} (1987) 2152.

\bibitem{Xu2003b}
R. X. Xu, {\it ApJ} {\bf 596} (2003) L59.

\bibitem{Zhou2004}
A. Z. Zhou, R. X. Xu, X. J. Wu, N. Wang, {\it Astroparticle Physics} {\bf 22} (2004) 73.

\bibitem{Xu2006}
R. X. Xu, D. J. Tao, Y. Yang, {\it MNRAS} {\bf 373} (2006) 85.

\bibitem{Xu2007}
R. X. Xu, {\it Advances in space research} {\bf 40} (2007) 1453.

\bibitem{Kouveliotou1998}
C. Kouveliotou, et al., {\it Nature} {\bf 393} (1998) 235.

\bibitem{Harding1999}
A. K. Harding, I. Contopoulos, D. Kazanas, {\it ApJ} {\bf 525} (1999) L125.

\bibitem{Ibrahim2002}
A. I. Ibrahim, et al., {\it ApJ} {\bf 574} (2002) L51.

\bibitem{Xu2003a}
R. X. Xu, H. G. Wang, G. J. Qiao, {\it Chinese Physics Letters} {\bf
20} (2003) 314.

\bibitem{GH07}
E.~V. Gotthelf, J.~P. Halpern, {\it ApJ} \textbf{664},
(2007) L35.

\bibitem{Xu2002}
R. X. Xu, {\it ApJ} {\bf 570} (2002) L65.

\bibitem{Katz1996}
J. I. Katz, {\it ApJ} {\bf 463} (1996) 305.

\bibitem{Zhang2000}
B. Zhang, R. X. Xu, G. J. Qiao, {\it ApJ}  {\bf 545} (2000) L127.

\bibitem{Spruit2010}
H. C. Spruit, (2010) arXiv:1005.5279.

\bibitem{Gavriil2008}
F. P. Gavriil, et al., {\it Science} {\bf 319} (2008) 1802.

\bibitem{Livingstone2010}
M. A. Livingstone, V. M. Kaspi, F. P. Gavriil, {\it ApJ} {\bf 710} (2010) 1710.

\bibitem{Rea2010}
N. Rea, et al., {\it Science} {\bf 330} (2010) 944.

\bibitem{Vink2006}
J. Vink, L. Kuiper, {\it MNRAS} {\bf 370} (2006) L14.

\bibitem{Gaensler2001}
B. M. Gaensler, P. O. Slane, E. V. Gotthelf, G. Vasisht, {\it ApJ} {\bf 559} (2001) 963.

\bibitem{Helfand2007}
D. J. Helfand, et al., {\it ApJ} (2007) {\bf 662} 1198.

\bibitem{ZhangBing2003}
B. Zhang, In: Stellar astrophysics - a tribute to Helmut A. Abt, {\it Astrophysics and Space Science Library}
{\bf 298} (2003) 27 (astro-ph/0212016).

\bibitem{Cheng2001}
K. S. Cheng, L. Zhang, {\it ApJ} {\bf 562} (2001) 918.

\bibitem{Mus2010}
S. Sasmaz Mus, E. Gogus, {\it ApJ} {\bf 723} (2010) 100.

\bibitem{Abdo2010}
A. A. Abdo, et al., {\it ApJ} {\bf 725} (2010) L73.

\bibitem{Tong2010}
H. Tong, L. M. Song, R. X. Xu, {\it ApJ} {\bf 725} (2010) L196.

\bibitem{Tong2011}
H. Tong, L. M. Song,R. X. Xu, {\it ApJ} {\bf 738} (2011) 31.

\bibitem{Parent2011}
D. Parent, et al., (2011) arXiv:1109.1590.

\bibitem{Turolla2011}
R. Turolla, et al., (2011) arXiv:1107.5488.

\bibitem{Muno2008}
M. Muno, et al., {\it ApJ} {\bf 680} (2008) 639.

\bibitem{Tong2010ins}
H. Tong, R. X. Xu, Q. H. Peng, L. M. Song, {\it Research in Astronomy and Astrophysics} {\bf 10} (2010) 553.

\bibitem{Tong2011ins}
H. Tong, R. X. Xu, L. M. Song, (2011) arXiv:1107.0830.

\bibitem{Levin2010}
L. Levin, et al., {\it ApJ} {\bf 721} (2010) L33.

\bibitem{Kaspi2007}
V. M. Kaspi, {\it ApSS} {\bf 308} (2007) 1.

\bibitem{Peng2007}
Q. H. Peng, H. Tong, {\it MNRAS} {\bf 378} (2007) 159.


\bibitem{Ouyed2011}
R. Ouyed, D. Leahy, B. Niebergal, {\it MNRAS} {\bf 415} (2011) 1590.

\bibitem{Malheiro2011}
M. Malheiro, J. A. Rueda, R. Ruffini, (2011) arXiv:1102.0653.

\bibitem{Frank2003}
J. Frank, A. King, D. Raine, Accretion power in
astrophysics, Cambridge University Press, Cambridge (2003).

\bibitem{Alpar2011}
M. A. Alpar, U. Ertan, S. Kaliskan, {\it ApJ} {\bf 732} (2011) L4.

\bibitem{Trumper2010}
J. E. Trumper, A. Zezas, U. Ertan, N. D. Kylafis, {\it A\&A} {\bf 518} (2010) 46.

\bibitem{Wang2006}
Z. X. Wang, D. Chakrabarty, D. L. Kaplan, {\it Nature} {\bf 440} (2006) 772.

\bibitem{Ertan2007}
U. Ertan, M. H. Erkut, K. Y. Eksi, M. A. Alpar, {\it ApJ} {\bf 657} (2007) 441.

\bibitem{Xu2011}
R. X. Xu, (2011) {\it IJMP} E, to appear (arXiv:1109.0665).

\bibitem{Chen2007}
A. B. Chen, T. H. Yu, R. X. Xu, {\it ApJ} {\bf 668} (2007) L55.

\bibitem{Paczynski2005}
B. Paczynski, P. Haensel, {\it MNRAS} {\bf 362} (2005) L4.

\bibitem{Cheng1998}
K. S. Cheng, Z. G. Dai, {\it PRL} {\bf 80} (1998) 18.

\bibitem{Usov2001}
V. V. Usov, {\it PRL} {\bf 87} (2001) 021101.

\bibitem{Lai2009}
X. Y. Lai, R. X. Xu, {\it Astroparticle Physics} {\bf 31} (2009) 128.

\bibitem{Kaspi2003}
V. M. Kaspi, F. P. Gavriil, P. M. Woods, {\it ApJ} {\bf 588} (2003) L93.

\bibitem{Helfand2001}
D. J. Helfand, E. V. Gotthelf, J. P. Halpern, {\it ApJ} {\bf 556} (2001), 380.

\end{thebibliography}
\end{document}